# BslA-stabilised emulsion droplets with designed microstructure

Keith M. Bromley[1] and Cait E. MacPhee[1,*]

[1] School of Physics and Astronomy, University of Edinburgh, James Clerk Maxwell Building, Peter Guthrie Tait Road, Edinburgh EH9 3FD, United Kingdom

Corresponding email address: cait.macphee@ed.ac.uk

**Abstract**

Emulsions are a central component of many modern formulations in food, pharmaceuticals, agrichemicals and personal care products. The droplets in these formulations are limited to being spherical as a consequence of the interfacial tension between the dispersed phase and continuous phase. The ability to control emulsion droplet morphology and stabilise non-spherical droplets would enable the modification of emulsion properties such as stability, substrate binding, delivery rate and rheology. One way of controlling droplet microstructure is to apply an elastic film around the droplet to prevent it from relaxing into a sphere. We have previously shown that BslA, an interfacial protein produced by the bacterial genus *Bacillus*, forms an elastic film when exposed to an oil- or air-water interface. Here, we highlight BslA's ability to stabilise anisotropic emulsion droplets. First, we show that BslA is capable of arresting dynamic emulsification processes leading to emulsions with variable morphologies depending on the conditions and emulsification technique applied. We then show that frozen emulsion droplets can be manipulated to induce partial coalescence. The structure of the partially coalesced droplets is retained after melting, but only when there is sufficient free BslA in the continuous phase. That the fidelity of replication can be tuned by adjusting the amount of free BslA in solution suggests that freezing BslA-stabilised droplets disrupts the BslA film. Finally, we utilise BslA's ability to preserve emulsion droplet structural integrity throughout the melting process to design emulsion droplets with a chosen shape and size.

## Introduction

Typically, liquid droplets adopt a spherical morphology in order to minimise their surface area and hence minimise their surface energy. However, this process can be arrested to give non-spherical droplet morphologies if an opposing elasticity is applied, either inside or at the surface of the droplet. For example, elasticity can be introduced internally by growing or encapsulating rigid filaments within emulsion droplets, as has been demonstrated in aqueous drops using actin filaments [1] and in oil droplets using wax crystals [2,3]. In the latter case, it was shown that anisotropic droplet morphologies could be arrested by the internal rigidity. Similarly, it is known that highly anisotropic morphologies result from arrested partial coalescence between solid fat droplets in ice cream [4]. *Interfacial* stabilisation of anisotropic liquid droplets and air bubbles has been achieved using a variety of different methods. Jamming of colloidal particles at interfaces has been shown to stabilise a range of non-spherical arrested morphologies including anisotropic emulsion droplets [5–7] and air bubbles [8], bicontinuous structures (bijels) [9], and recently water-in-air "liquid plasticine" [10]. Anisotropic droplets have also been created using systems that do not utilise colloidal particles. For example, faceted droplets and droplets with tails have been formed via interfacial freezing of hexadecane and a surfactant; in this system the interfacial elastic energy could overcome the near-zero interfacial tension [11,12]. Proteins have also been used to stabilise anisotropic structures. Specifically, a class of fungal proteins called hydrophobins were able to stabilise anisotropic air bubbles [13] and emulsion droplets [14], as well as being able to deform sessile drops of water in air [15].

BslA is a bacterial protein from *Bacillus subtilis* [16–19] that is similar in function to the fungal hydrophobins, although it is distinct in terms of structure, sequence and mechanism of action [20,21]. *In vitro* studies elucidated that the purified protein is surface active due to a conformational change that occurs only once the protein encounters a hydrophobic interface [21]. Further, we have previously shown that BslA can form an elastic film at both oil-water and air-water interfaces [20,21] and that the elastic film can stabilise anisotropic air bubbles [22]. Such qualities make BslA an excellent candidate for stabilising emulsions. Indeed, the study that demonstrated the conformational change at the oil-water interface utilised BslA-stabilised emulsions for that purpose [21], and functionalised emulsion "microcapsules" have been prepared using BslA as the stabiliser [23].

Here, we have created anisotropic liquid emulsion droplets with an elastic BslA surface. Initially, we demonstrate BslA's ability to arrest dynamic emulsification processes, such as droplet elongation, breakup and coalescence. Thus, by varying emulsification conditions, we can broadly control the resulting droplet morphology. Cooling the dispersed phase to solidify the droplets and then partially coalescing them achieved further control over emulsion microstructure. Such droplets could retain their partially coalesced structures during and after melting. Finally, to exert control over the final emulsion architecture, we created moulded fat droplets by casting them in a cylindrical template. The moulded fat droplets were released into a cold BslA solution and subsequently warmed to melt the internal fat phase. Despite melting the fat, the development of an interfacial elastic BslA film allowed the droplets to retain their original moulded morphology.

# Experimental

BslA preparation

The BslA used throughout this research was a form of BslA in which cysteine amino acids were replaced with alanine, to prevent the potential formation of disulphide-bonded oligomers. To achieve this, the two cysteine residues at positions 178 and 180 were replaced by site-specific mutagenesis, creating a C178,180A mutant (Nicola Stanley-Wall, unpublished). The protein was expressed and purified as described in [21]. All BslA solutions were prepared in 25 mM pH 7 phosphate buffer that had been filtered through a 0.22 µm filter.

Emulsion preparation

Emulsification of hexadecane into BslA solution was performed using three traditional methods: high shear mixing; vortex mixing; and probe sonication. For high shear mixing, an Ultra-Turrax T10 rotor-stator with a gap width of 0.3 mm and rotor speed of 30000 rpm was used. This corresponds to a shear rate of 20000 s$^{-1}$. Samples were prepared at a range of BslA concentrations (0.05 – 0.5 mg/mL) and oil volume fractions ($\phi_o$) (0.01 – 0.5) and emulsified for 15 s unless specified. Vortex mixing was performed for 60 s using an IKA Vortex Genius 3 operating at 2500 rpm. Probe sonication was performed using a Sonics Vibra-Cell VCX-500 probe sonicator equipped with a 3 mm tapered titanium microtip. Sonication was performed in 1 s pulses (amplitude = 20%) with 5 s pauses between each pulse for a total sonication time of 30 s. Emulsions created using vortex mixing or sonication were prepared at a BslA concentration of 0.2 mg/mL and at $\phi_o = 0.2$.

Two alternative emulsification methods were also performed: a rollerbank method and a "floccing" method. In the rollerbank method, a vial was completely filled with 50% 0.2 mg/mL BslA solution/50% hexadecane (v/v). This vial was then placed on a Stuart SRT9 roller mixer rotating at 36 rpm for 24 hours. In the floccing method, a 200 µL pipette tip was filled with 200 µL hexadecane and then transferred to a pipettor set to dispense 100 µL. 100 µL of the hexadecane was then dispensed from the tip and replaced with 100 µL 0.2 mg/mL BslA. The BslA solution was then dispensed and refilled from the tip by rapidly manually aspirating at least ten times to produce an emulsion.

Partial coalescence of frozen emulsion droplets was achieved by centrifuging emulsions at 17000 $g$ in a Thermo Scientific Heraeus Fresco 21 microcentrifuge. Centrifugation was performed for 5 minutes at 5 ºC. In samples prepared for confocal laser scanning microscopy, the hexadecane phase contained 100 µM Nile Red.

Emulsion droplet moulding

Cylindrical droplets with a defined diameter of ~210 µm were produced by using a 27G needle as a template. The entire templating process was performed in a cold room at 5 °C. First, a vial of coconut oil was cooled to 5 °C until it was completely frozen. Then a 27G needle was pushed into the frozen coconut oil. When the needle was removed, a cylinder of frozen coconut oil remained inside. This cylinder was physically ejected (using a copper wire) directly into cold BslA solution (0.2 mg/mL) that had been pipetted into the cavity of a microscope slide. Once several cylinders had been produced, the sample was covered with a 22 x 50 mm cover slip and imaged under a microscope. After imaging, the sample was warmed to 30 °C and the droplets imaged again.

Laser diffraction particle size analysis

Emulsion droplet sizing was performed using a Beckman-Coulter LS 13 320 particle size analyser. Emulsion samples were pipetted into 25 mM pH 7 phosphate buffer (stirred) in a cuvette until an obscuration value of ~10% was reached. Samples were measured at least three times with a measurement time of 60 s and the average size distribution was recorded. Particle size distributions were converted to the Sauter mean diameter, $d_{32}$, also known as the volume-surface diameter as it represents the droplet size that reflects the volume:surface area ratio of the entire system.

Optical microscopy

Emulsion droplet morphologies were analysed using an Olympus BX-50 microscope equipped with a QImaging QICAM digital camera. Emulsions were diluted to $\phi_o$ = 0.001 to 0.005 depending on the size of the droplets and pipetted into the cavity of a microscope slide. A 22 x 50 mm cover slip was immediately placed over the top of the droplet to allow the droplets to cream onto the cover slip surface. In some experiments, the emulsion was pipetted directly into buffer solution in a cavity half-covered by a cover slip. This avoided elongation of large droplets at the air-water interface due to the application of the cover slip.

Confocal laser scanning microscopy

Hexadecane emulsion samples were prepared containing 100 µM Nile Red dye. An Instec TSA02i temperature stage was used to adjust the temperature during melting experiments. A 7 µL drop of emulsion at $\phi_o$ = 0.01 was placed on a 22 x 50 mm cover slip with 45 µm thick tape used as a spacer. The sample was then covered with an 18 x 18 mm cover slip. In temperature controlled experiments, slide assembly and sample mounting was performed *in situ*. A Zeiss LSM 700 confocal laser-scanning microscope was used to image the fluorescently labelled emulsion droplets. Samples were excited with a 10mW, 555 nm solid state laser attenuated to 10% intensity.

# Results and Discussion

Emulsions prepared using BslA as a stabiliser generally formed a mixture of spherical and aspherical droplets. The morphology of the aspherical droplets varied depending on the conditions and emulsification method used.

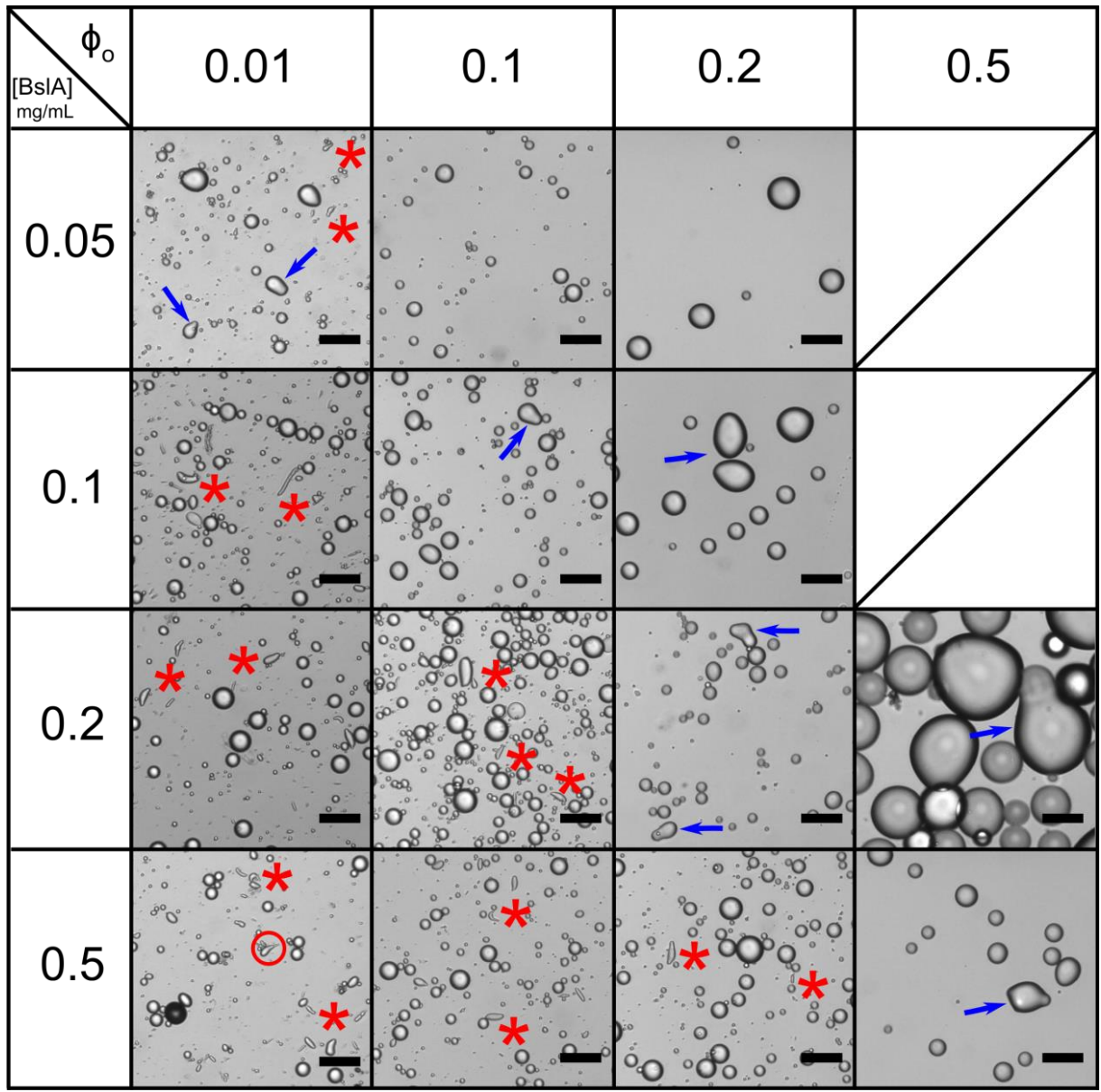

*Figure 1: Emulsions prepared at different BslA concentration and oil volume fraction conditions by mixing in a high shear mixer for 15s. The morphologies of the non-spherical droplets changes as the system changes from the emulsifier-rich to the emulsifier-poor regime. Under emulsifier-rich conditions, rod-shaped droplets are stabilised (indicated by red asterisks). In the bottom-left panel, a teardrop structure is circle in red. Snowman droplets (indicated by blue arrows) are not observed under emulsifier-rich conditions, but appear under emulsifier-poor conditions. Rod-shaped droplets are not observed under emulsifier-poor conditions. The blank panels denote that a stable emulsion was not formed. All scale bars = 50 μm.*

Hexadecane emulsions were prepared by high-shear mixing at [BslA] = 0.05, 0.1, 0.2 and 0.5 mg/mL and $\phi_o$ = 0.01, 0.1, 0.2 and 0.5. Figure 1 shows representative images of the emulsions and Figure 2 shows particle sizing data for each preparation condition. At the lowest oil volume fraction $\phi_o$ = 0.01, all four emulsion samples contained rod (red asterisks) and teardrop shaped droplets, while partially coalesced "snowman" shaped droplets were also seen at the lowest

BslA concentration (0.05 mg/mL; blue arrows). At this oil volume fraction $d_{32}$ increased slightly with increasing BslA concentration, however we also observed the formation and stabilisation of a foam when mixing high concentrations of BslA at low volume fractions of oil, thus BslA-stabilised air bubbles in the emulsion are likely to have skewed the data.

On increasing the hexadecane content to $\phi_o = 0.1$, rod-shaped droplets were the dominant anisotropic droplet at [BslA] = 0.2 and 0.5 mg/mL. At 0.1 mg/mL protein, rods were no longer observed, but "snowman" morphologies were present. Only spherical droplets were observed at [BslA] = 0.05 mg/mL and these were larger ($d_{32} > 20$ µm) than the emulsions prepared at the same oil volume fraction but higher protein concentration.

At $\phi_o = 0.2$, rod-shaped droplets were only observed at the highest BslA concentration. At [BslA] = 0.2 mg/mL snowman droplets could be seen. At [BslA] = 0.1 mg/mL, most of the aspherical droplets appeared ovoid in shape. Only spherical droplets were observed at [BslA] = 0.05 mg/mL. The $d_{32}$ value decreased considerably with increasing BslA concentration from ~50 µm at 0.05 mg/mL to ~15 µm at 0.2 and 0.5 mg/mL.

Only the two higher BslA concentrations could stabilise emulsions at $\phi_o = 0.5$. Both emulsions contained snowman morphologies and no rods, while $d_{32}$ was far greater than observed for those BslA concentrations at lower $\phi_o$.

From Figure 1, it is clear that, alongside spherical emulsion particles, two types of anisotropic droplet are typically stabilised during emulsification by high shear mixing – rods or snowmen. Teardrops are also sometimes observed alongside rods (Figure 1, bottom left panel). Rods are stabilised when the BslA:oil ratio is high enough. Under conditions where less BslA is available for adsorption, either due to a lower BslA concentration or increased $\phi_o$, rods are no longer observed and snowman structures are instead seen. These observations are consistent with predicted emulsification mechanisms under "emulsifier-rich" and "emulsifier-poor" conditions [24].

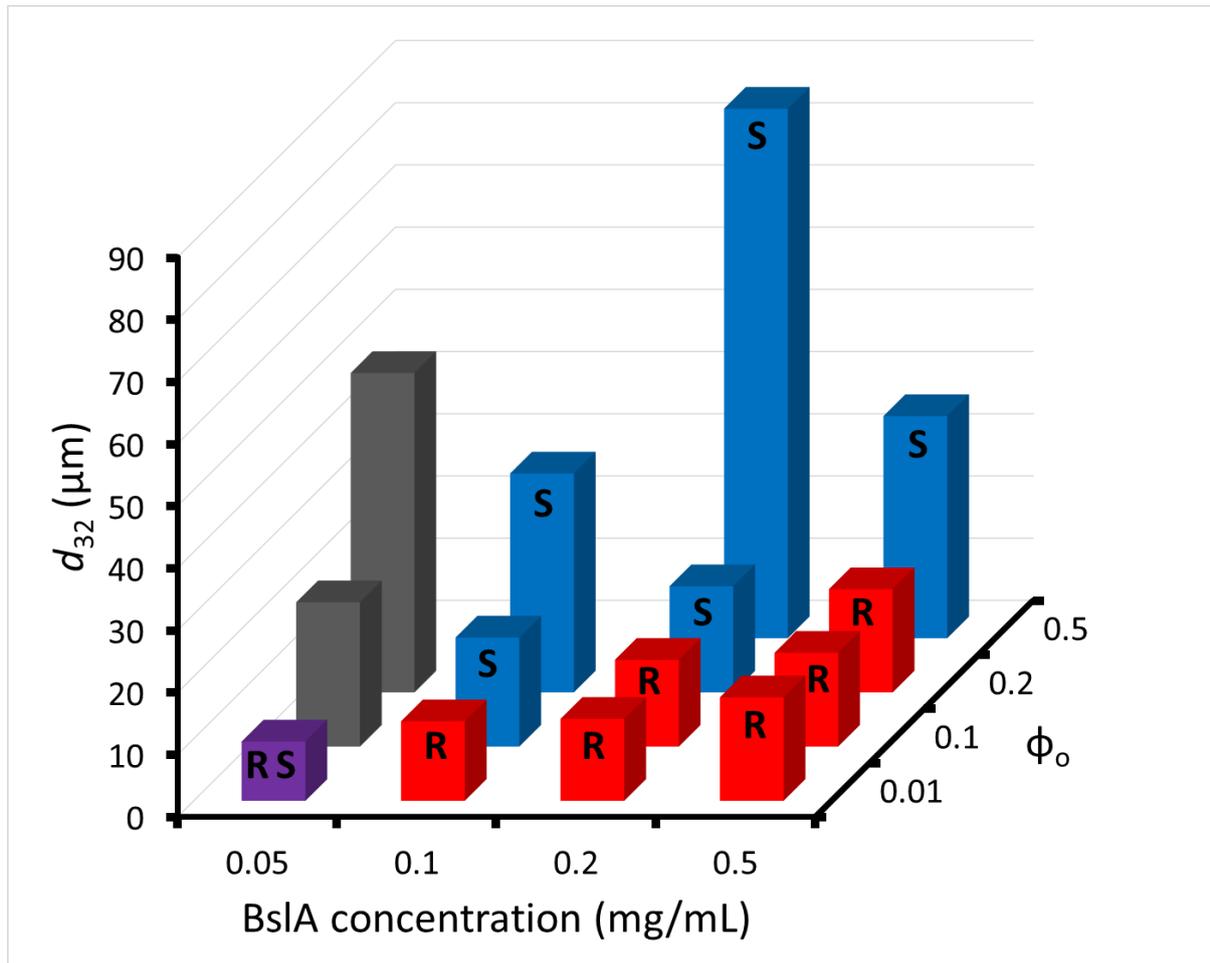

*Figure 2: Laser diffraction particle size analysis of emulsions prepared at different BslA concentration and oil volume fraction conditions by mixing in a high shear mixer for 15s. The colour of and the letters on the bars indicate the main anisotropic droplet morphology that was observed in each emulsion: rods (red, R), snowmen (blue, S), both morphologies (purple, RS) and neither morphology (grey).*

In the emulsifier-rich regime, there is sufficient emulsifier to stabilise droplets during droplet breakup such that the smallest droplets achievable under the chosen hydrodynamic conditions are fully coated in protein. When a droplet splits in two, a small filament develops as the droplet is stretched. The filament ultimately breaks apart and the resultant droplets that form from the pinched-off filament are usually identified as "satellite droplets" [25,26]. In the presence of sufficient BslA, these filaments are stabilised. The presence of other anisotropic drop shapes such as "teardrops" and larger elongated drops (Figure 1, bottom left panel) supports the idea that the rods are indeed a product of the droplet breakup process as remnants of the "mother drop" would necessarily be formed with the satellite drops. Importantly, as the products of droplet breakup quickly develop a full monolayer of BslA, partial coalescence of colliding droplets is inhibited and thus snowman structures are not usually observed.

In the emulsifier-poor regime, there is no longer sufficient BslA present to stabilise the products of droplet breakup, so rods and teardrops are no longer observed under high shear mixing. High shear mixing creates high and low perturbation zones during the emulsification process [27]. In the high shear zone the shear rate is 20000 s$^{-1}$, which corresponds to a timescale of droplet deformation and breakup of roughly 50 µs [28]. This short droplet deformation and breakup time decreases the probability that BslA could stabilise the transient structures formed via droplet breakup under emulsifier-poor conditions. However, in such emulsifier-limiting conditions, many of the droplets would be partially coated [24] as they emerge from the high shear zone. When two droplets *partially* coated in BslA collide, the coalescence is quickly arrested by the BslA layer and a snowman droplet is created. The Smoluchowski expression for collision rate in laminar flow [29] estimates that each droplet would experience roughly 5000 collisions per second (for droplet diameter = 16 µm, $\phi_o$ = 0.2, shear rate, $\dot{\gamma}$ = 20000 s$^{-1}$) within the high shear zone. Given the dimensions of the rotor-stator head, the high shear zone in our system accounts for roughly 5% of the total emulsion volume. If the majority of collisions resulted in partial coalescence, we would expect to observe a much higher yield of snowmen. The low yield of snowmen droplets suggests that partial coalescence can only occur over a limited range of surface fractions of colliding droplets. Indeed, this surface fraction range has been previously estimated to be between ~0.7 and 0.9 for droplets colliding in a particle-stabilised emulsion system [5]. Below an average surface fraction of 0.7, coalescence could not be arrested and above a surface fraction of 0.9 on an individual droplet, coalescence would not occur. Given the low yield of snowmen, it seems likely that colliding BslA-stabilised droplets have similar limits to the arrest of coalescence as particle stabilised emulsions [2,5,30]. At extremely low BslA concentration, the partially coated droplets have so little BslA adsorbed at the surface that the snowman morphology can no longer be arrested. This lower limit of anisotropic shape stabilisation occurs at roughly 0.05 mg/mL for $\phi_o = 0.1 - 0.2$ (Figure 1).

When emulsions were prepared via sonication, the predominant aspherical droplets formed also had the snowman morphology (Figure 3a). There are two major parallels between the high shear mixing technique and emulsification by probe tip sonication. Firstly, a high intensity cavitation zone develops around the probe tip, leaving a "dead zone" closer to the edge of the sonication vessel [31], much like the high and low shear zones that exist in a high shear mixer. Secondly, the timescale of cavitation bubble collapse is on the order of less than 1 µs [32], which is even shorter than the timescale of droplet deformation and breakup during high shear mixing (50 µs for a shear rate of 20000 s$^{-1}$ [28]). As such, under the same emulsifier-poor conditions, it is not surprising that droplet breakup cannot be stabilised during sonication and thus a similar droplet morphology is produced from both sonication and high-shear mixing (Figure 3a, b).

The emulsifier-poor conditions that produced snowman droplets in the high shear mixer ([BslA] = 0.2 mg/mL and $\phi_o = 0.2$) could not stabilise rods as the droplet breakup process occurred too rapidly for the low concentration of available protein to stabilise. If the droplet breakup process was slowed by changing the emulsification method, could rods be stabilised under the same emulsifier-poor conditions?

By lowering the shear rate from 20000 s$^{-1}$ to 5000 s$^{-1}$, emulsions prepared at [BslA] = 0.2 mg/mL and $\phi_o = 0.2$ had rod and teardrop morphologies (Figure 3c) instead of the snowman morphology that was prevalent at the higher shear rate.

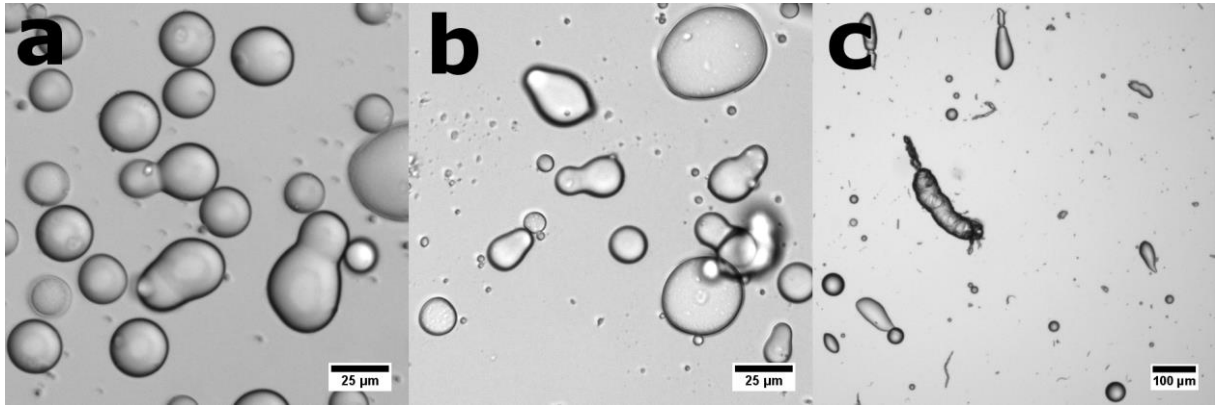

*Figure 3: (a) Hexadecane snowman droplets formed via emulsification using a probe sonicator (b) Hexadecane snowman droplets formed via emulsification using a high-shear mixer at a shear rate of 20000 $s^{-1}$. (c) Hexadecane teardrops and rods formed via emulsification using a high shear mixer at a shear rate of 5000 $s^{-1}$. All three emulsions (a) – (c) were prepared at [BslA] = 0.2 mg/mL and $\phi_o = 0.2$.*

Equally, preparing emulsions by shaking on a vortex mixer created long rod-shaped emulsion droplets with aspect ratios as high as 50 (Figure 4a). The widths of the rod-shaped droplet in these samples ranged from approximately 1 to 10 µm. As with rods formed under high shear mixing, teardrop shaped emulsions were observed alongside the rod-shaped droplets (Figure 4a). High aspect ratio droplets were also formed by slowly rolling a vial containing oil and BslA solution on a rollerbank for 24 hours. In that case, only a small percentage of the oil was emulsified, but the droplets that formed were extremely small, with the lateral diameter of the rods often measuring below 1 µm (Figure 4b). Confocal laser-scanning microscopy of the droplets stained with Nile Red established that they were indeed oil droplets (Figure 4c) and not BslA stabilised rod-shaped air bubbles as have been observed previously [22]. Repeatedly shearing BslA in the presence of hexadecane in a pipettor tip (the floccing method, see Experimental) produced rod-shaped droplets of a similar size and aspect ratio (Figure 4d) to the droplets created by the rollerbank method.

The stabilisation of rods by vortex mixing under emulsifier-poor conditions ([BslA] = 0.2 mg/mL, $\phi_o = 0.2$) (Figure 4a) can be explained by considering the shear profile in the system.

Vortex mixers have a relatively large shear zone that leads to long residence times [27] and a lower shear rate compared to high shear mixers. Those differences result in a slower droplet breakup process as the timescale of droplet deformation is inversely proportional to the shear rate [28]. We suggest that in this case the droplet breakup process is slowed enough for BslA to be able to stabilise the filament and teardrops that develop as a droplet breaks apart even when in the emulsifier-poor regime.

Rods are also the predominant anisotropic droplet formed via the rollerbank method (Figure 4b, c). Typically, the production of extremely fine emulsion droplet sizes with diameters on the order of a few microns requires the entire bulk oil phase (for oil-in-water emulsions) to be broken up and suspended in the aqueous phase, a process that requires high energies. The rollerbank process is, in contrast, a low energy method that produced fine oil droplets directly from the bulk phase without breaking down the bulk oil phase entirely. We hypothesise that the ability to generate emulsion droplets by gently rotating separate hexadecane and aqueous BslA phases at a 1:1 ratio on a rollerbank is the result of variations in shear stress within the BslA nanofilm formed at the oil-water interface. In regions of increasing shear stress such as when the BslA nanofilm approaches the vial wall, the BslA nanofilm would fracture as the shear stress applied to the interface would result in extensile stress *within* the interface. Fracturing would expose fresh oil-water interface for excess BslA in the aqueous phase to adsorb onto, thus increasing the total surface area of interfacially bound BslA. The increase in interfacial surface area would ultimately lead to the formation of wrinkles at the interface, which we suggest close in on themselves and bleb off the interface as rod-shaped droplets. This mechanistic idea was supported by the observation that the floccing method, a different way of applying shear stress, also produced small rod-shaped droplets with very similar size and morphology to the droplets generated by the rollerbank method (Figure 4d). In this case, a BslA nanofilm could form along the inside of the pipettor tip, which would be wet by hexadecane.

As we hypothesized for the rollerbank method, the BslA nanofilm at the oil-water interface within the pipette tip would experience similar extensile and compressive stresses during floccing, except in this case the stress applied can be controlled by the rate of aspiration. Similar observations have been made by shearing BslA at an air-water interface, leading to the formation of a wide range of higher order BslA assemblies [19,22].

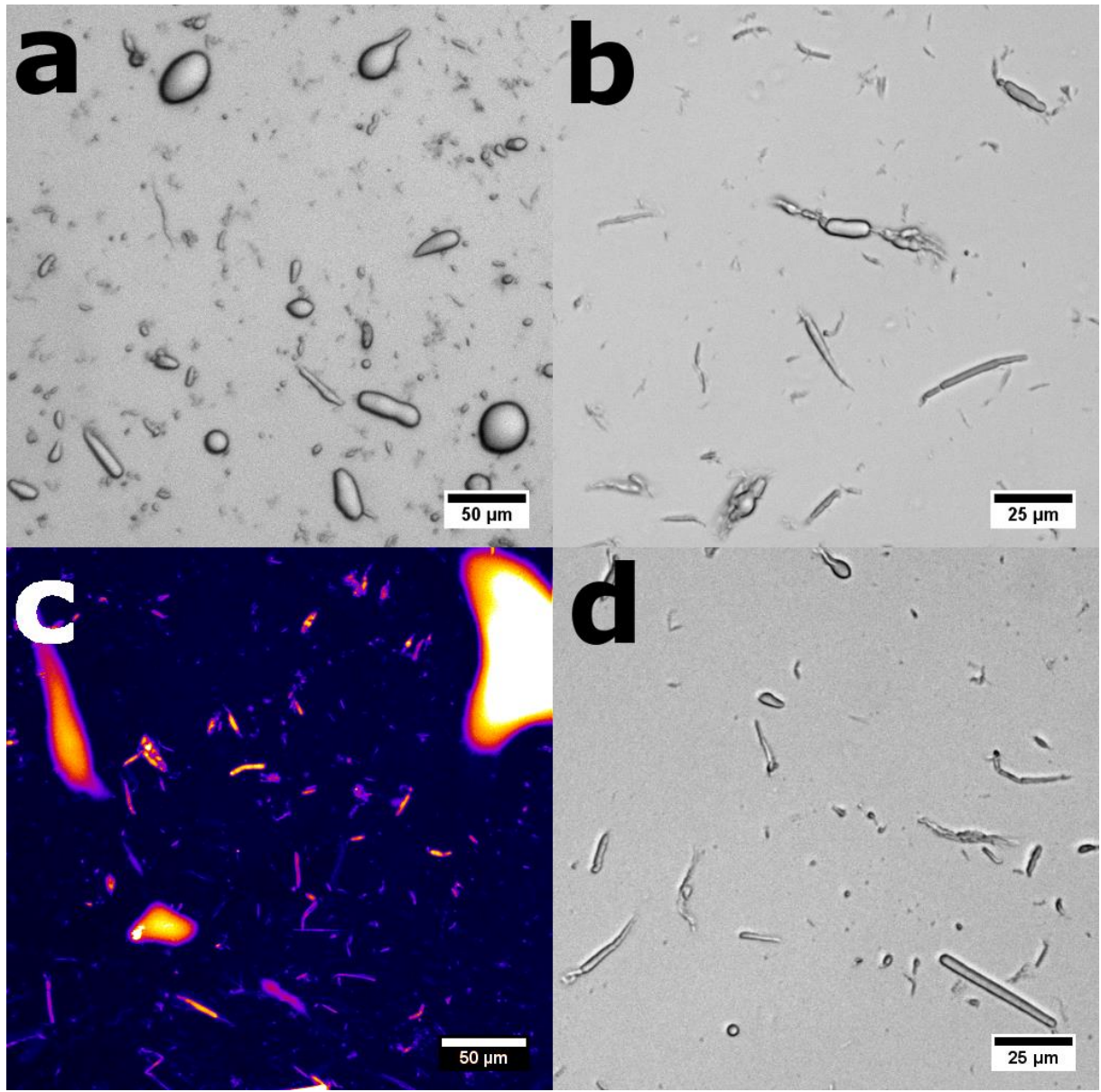

*Figure 4: Images of emulsion droplets arrested during emulsification. (a) Elongated droplets trapped during vortex mixing at [BslA] = 0.2 mg/mL and $\phi_o$ = 0.2. Note the presence of a teardrop-shaped droplet near the top of the image. (b) Rod-shaped droplets formed by rolling a vial of BslA and hexadecane on a rollerbank for 24 hours. (c) Rod-shaped rollerbank droplets stained with Nile Red and viewed using confocal laser-scanning microscopy. (d) Rod-shaped oil droplets created by "floccing" BslA solution in a pipettor tip in the presence of hexadecane.*

Although they were the predominant type of non-spherical droplet observed in emulsifier-poor conditions, the overall yield of snowman structures formed by sonication and high-shear mixing was very low. By centrifuging BslA-stabilised hexadecane droplets, we could increase the incidence of droplet-droplet collisions, thus promoting the creation of snowman droplets. This was true whether or not we froze the hexadecane emulsion prior to centrifugation. Using emulsions that were frozen and then centrifuged, we observed the melting process to determine how efficiently BslA could maintain the shape of the partially coalesced droplets. Figure 5a shows a triplet of hexadecane droplets melting in a cleaned system where excess BslA had been removed. Although the overall "Mickey Mouse" morphology of the triplet was maintained, significant relaxation at the joints could be observed. Relaxation at the joints was a common feature of melting doublets and triplets in systems with no excess BslA. This meant the final surface area of the melted doublets and triplets was significantly lower than it would be for two spherical oil droplets (with an equivalent total volume to the doublet or triplet) conjoined by a small neck. The relaxation of the neck suggests that the freeze-thaw process disrupts the interfacial BslA nanofilm. We propose that this is a consequence of the formation of planar crystalline surfaces at the frozen hexadecane droplet interface, resulting in a higher relative surface area per droplet. With no excess BslA available to adsorb onto the newly exposed hexadecane interface, the broken BslA nanofilm requires time on the order of a few seconds [21] to reorganise during the melting process. The overall decrease in droplet surface area implies that some BslA is lost from the interface during a freeze-thaw event. To test the idea that neck relaxation was a consequence of BslA reorganisation and potentially loss of BslA during the freeze-thaw process, the experiment was repeated with excess BslA in the aqueous phase. As expected, far less relaxation was observed at the necks (Figure 5b) demonstrating that although some rearrangement and/or loss of interfacially bound BslA still occurred, the free BslA in the aqueous phase could adsorb onto the bare hexadecane surfaces to repair the ruptured nanofilm. Ultimately, the availability of free BslA in the aqueous phase allowed for a more faithful replication of the original frozen morphology.

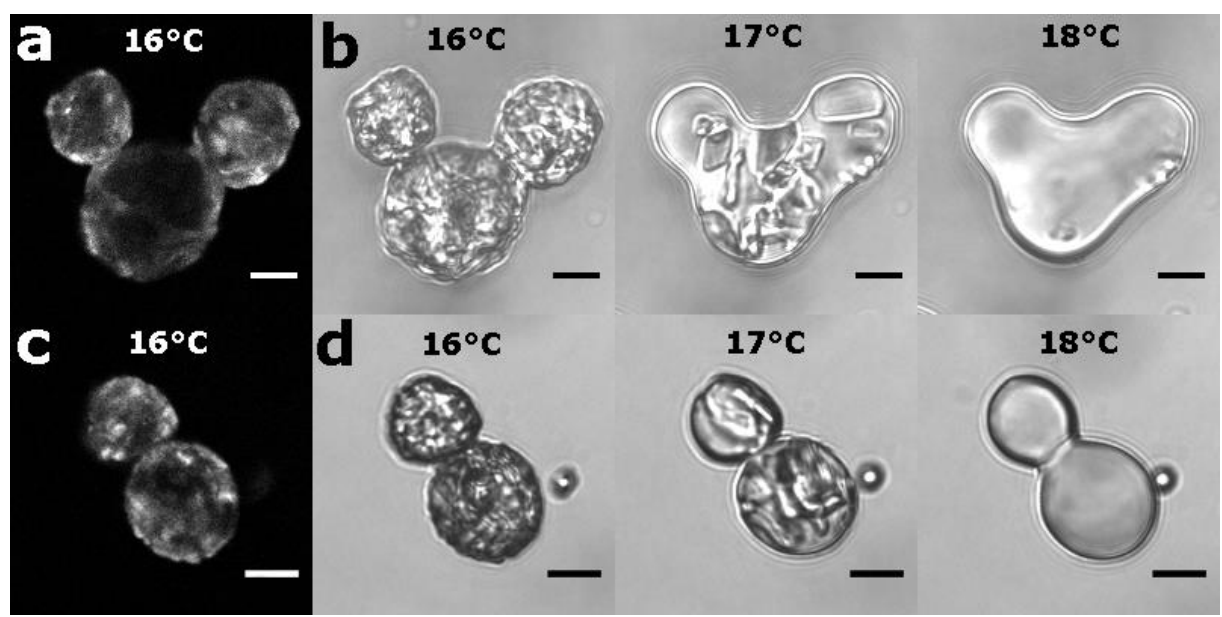

*Figure 5: Melting transitions of BslA stabilised partially coalesced frozen hexadecane droplets with (a) and (b) no excess BslA and (c) and (d) 0.9 mg/mL BslA in the continuous phase. (a) and (c) are confocal images of the frozen droplets stained with Nile Red. The image sequences (b) and (d) are transmission images taken before melting at ~16 °C, during melting at ~17 °C, and after melting at ~18 °C. Relaxation at the neck occurred more significantly in the absence of free BslA. The ramp rate was 0.5 °C/min. All scale bars = 10 μm.*

In the previous method, we demonstrated that BslA could preserve anisotropic droplet shapes while the internal oil phase was melted. However, the anisotropic snowman droplets were based on the coalescence of two or three preformed spherical droplets. Can this method of interfacial stabilisation allow us to design a method that can produce droplets with specific morphologies and dimensions? Instead of creating an emulsion using one of the traditional methods, we used a 27G needle with an inner diameter of ~210 μm to template a core of fat (cooled coconut oil, $T_m \approx 25°C$) that was extruded directly into a cold BslA solution ($T < T_m$ of the oil phase), thus negating any driving force towards the formation of spherical droplets. Upon melting the fat by warming to 30°C, the morphology of the coconut oil cylinders remained virtually identical to that of the original fat cylinders (Figure 6) demonstrating that BslA had successfully formed an elastic interface around the large rod-shaped oil droplet. As observed with the melting snowman morphologies in the presence of excess BslA, a very small amount of relaxation occurred.

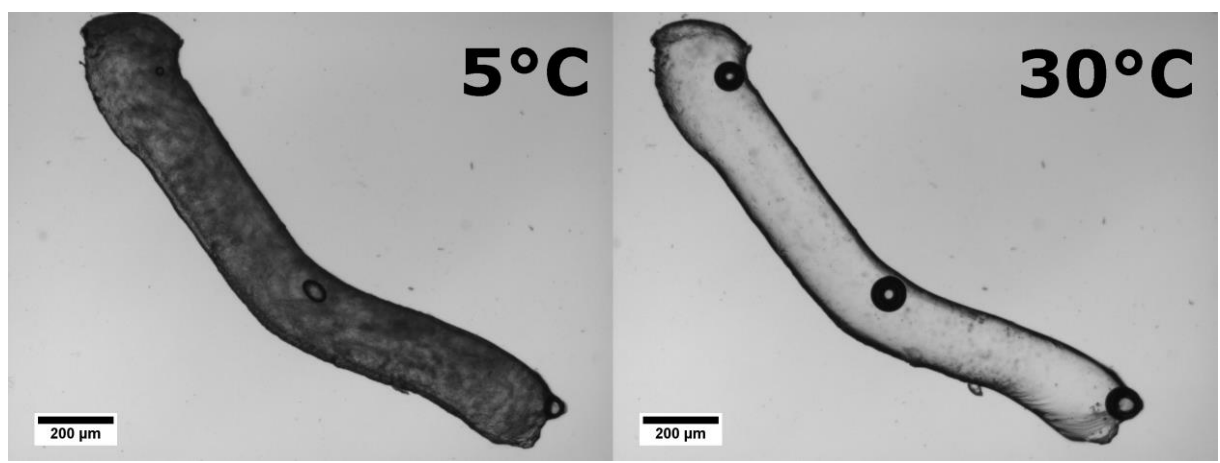

*Figure 6: Left, coconut fat rod-shaped emulsion drop formed by extruding the fat from a 27G needle into BslA solution at 5 °C. Right, the same droplet after warming to 30 °C. Even after the fat had melted to oil, the overall morphology of the droplet was maintained by the elastic BslA film at the interface.*

We have demonstrated that the interfacial protein BslA can be used to stabilise emulsions with programmed size and shape. This work could be developed to produce highly monodisperse anisotropic emulsions using microfluidics in an extension of the work of Xu et al., which has shown that monodisperse cylindrical and discoid solid particles can be created by thermal setting within a microfluidic device [33]. A similar microfluidic method has recently been exploited to create monodisperse anisotropic droplets stabilised by an internal crystal endoskeleton [34]. Using a 5 nm thick BslA layer to stabilise these emulsions offers several advantages over alternative systems. Firstly, the emulsions do not require a solid scaffold to retain their anisotropic morphology, as is the case if anisotropic emulsions are stabilised using internal fat crystals [2,3], allowing the preparation of entirely liquid anisotropic emulsions with only a very thin protein sheet as a stabiliser. Secondly, the solid templating technique that we used to generate designed emulsion droplets is only possible if the film-forming stabiliser can either adsorb onto solid surfaces or adsorb onto a melting interface quickly enough to arrest droplet relaxation. The technique enables the creation of liquid emulsion droplets that are extremely close replicas of the originally cast solid shape. This is an improvement over a previous method of extruding liquid oil through capillaries that created elongated droplets with widths broadly similar to the capillary width [7], which relied on particles adsorbing to a liquid-liquid interface. Thirdly, BslA has a rich surface chemistry that should enable facile modification of the surface of the emulsions and proteins offer the advantage of being genetically modifiable to allow specific functionalisation of the emulsion surface. This has already been demonstrated to create fluorescently labelled BslA-stabilised emulsions [23].

This control over droplet morphology should lead to the development of methods to produce emulsions with defined microstructure in the future. However, production of commercial formulations generally requires high throughput methods and low-throughput templating methods may be unachievable in an industrial setting, so developing an understanding of how BslA influences droplet structure during dynamic emulsification processes is also important. Here, we have demonstrated how and why different emulsification methods produce different emulsion morphologies. With this knowledge, it should be possible to design protocols to produce emulsion populations with uniform microstructure. For example, by applying low shear in a flow-through system containing BslA and an oil, it should be possible to create almost exclusively rod-shaped emulsions in large batches. With such control over droplet morphology attainable through a variety of processes, incorporating BslA into functional emulsion formulations looks like a realistic prospect.

## Competing interests

We have no competing interests.

## Authors' contributions

KMB designed and performed experiments, drafted the manuscript

CEM conceived of the study, coordinated the study and prepared the manuscript

**Funding**

EPSRC grant no. EP/J007404/1

BBSRC grant no. BB/M013774/1

**Acknowledgements**

We would like to thank Professor Nicola Stanley-Wall and Dr Tetyana Sukhodub of the University of Dundee for creating and providing the modified BslA protein used in this work.

**References**

1. Claessens, M. M. A. E., Tharmann, R., Kroy, K. & Bausch, A. R. 2006 Microstructure and viscoelasticity of confined semiflexible polymer networks. *Nat. Phys.* **2**, 186–189. (doi:10.1038/nphys241)

2. Pawar, A. B., Caggioni, M., Hartel, R. W. & Spicer, P. T. 2012 Arrested coalescence of viscoelastic droplets with internal microstructure. *Faraday Discuss.* **158**, 341–350. (doi:10.1039/c2fd20029e)

3. Caggioni, M., Bayles, A. V, Lenis, J., Furst, E. M. & Spicer, P. T. 2014 Interfacial stability and shape change of anisotropic endoskeleton droplets. *Soft Matter* **10**, 7647–7652. (doi:10.1039/C4SM01482K)

4. Clarke, C. 2004 *The Science of Ice Cream*. 1st edn. Cambridge: The Royal Society of Chemistry.

5. Pawar, A. B., Caggioni, M., Ergun, R., Hartel, R. W. & Spicer, P. T. 2011 Arrested coalescence in Pickering emulsions. *Soft Matter* **7**, 7710–7716. (doi:10.1039/c1sm05457k)

6. Clegg, P. S., Herzig, E. M., Schofield, A. B., Horozov, T. S., Binks, B. P., Cates, M. E. & Poon, W. C. K. 2005 Colloid-stabilized emulsions: behaviour as the interfacial tension is reduced. *Langmuir* **1**, 1–6. (doi:10.1088/0953-8984/17/45/031)

7. Bon, S. A. F., Mookhoek, S. D., Colver, P. J., Fischer, H. R. & van der Zwaag, S. 2007 Route to stable non-spherical emulsion droplets. *Eur. Polym. J.* **43**, 4839–4842. (doi:10.1016/j.eurpolymj.2007.09.001)

8. Bala Subramaniam, A., Abkarian, M., Mahadevan, L. & Stone, H. A. 2005 Colloid science: Non-spherical bubbles. *Nature* **438**, 930. (doi:10.1038/438930a)

9. Herzig, E. M., White, K. A., Schofield, A. B., Poon, W. C. K. & Clegg, P. S. 2007 Bicontinuous emulsions stabilized solely by colloidal particles. *Nat. Mater.* **6**, 966–971. (doi:10.1038/nmat2055)

10. Li, X., Xue, Y., Lv, P., Lin, H., Du, F., Hu, Y., Shen, J. & Duan, H. 2015 Liquid plasticine: controlled deformation and recovery of droplet with interfacial nanoparticle jamming. *Soft Matter* **12**, 1655–1662. (doi:10.1039/C5SM02765A)

11. Denkov, N., Tcholakova, S., Lesov, I., Cholakova, D. & Smoukov, S. K. 2015 Self-shaping of oil droplets via the formation of intermediate rotator phases upon cooling. *Nature* **528**, 392–395. (doi:10.1038/nature16189)

12. Guttman, S., Sapir, Z., Schultz, M., Butenko, A. V., Ocko, B. M., Deutsch, M. & Sloutskin, E. 2016 How faceted liquid droplets grow tails. *Proc. Natl. Acad. Sci.* **113**, 493–496. (doi:10.1073/pnas.1515614113)

13. Basheva, E. S., Kralchevsky, P. A., Christov, N. C., Danov, K. D., Stoyanov, S. D., Blijdenstein, T. B. J., Kim, H. J., Pelan, E. G. & Lips, A. 2011 Unique properties of bubbles and foam films stabilized by HFBII hydrophobin. *Langmuir* **27**, 2382–2392.

(doi:10.1021/la104726w)

14. Rosu, C., Kleinhenz, N., Choi, D., Tassone, C. J., Zhang, X., Park, J. O., Srinivasarao, M., Russo, P. S. & Reichmanis, E. 2016 Protein-Assisted Assembly of π-Conjugated Polymers. *Chem. Mater.* **28**, 573–582. (doi:10.1021/acs.chemmater.5b04192)

15. Szilvay, G. R., Paananen, A., Laurikainen, K., Vuorimaa, E., Lemmetyinen, H., Peltonen, J. & Linder, M. B. 2007 Self-assembled hydrophobin protein films at the air-water interface: structural analysis and molecular engineering. *Biochemistry* **46**, 2345–2354. (doi:10.1021/bi602358h)

16. Kearns, D. B., Chu, F., Branda, S. S., Kolter, R. & Losick, R. 2005 A master regulator for biofilm formation by Bacillus subtilis. *Mol. Microbiol.* **55**, 739–749. (doi:10.1111/j.1365-2958.2004.04440.x)

17. Branda, S. S., Chu, F., Kearns, D. B., Losick, R. & Kolter, R. 2006 A major protein component of the Bacillus subtilis biofilm matrix. *Mol. Microbiol.* **59**, 1229–1238. (doi:10.1111/j.1365-2958.2005.05020.x)

18. Ostrowski, A., Mehert, A., Prescott, A., Kiley, T. B. & Stanley-Wall, N. R. 2011 YuaB functions synergistically with the exopolysaccharide and TasA amyloid fibers to allow biofilm formation by Bacillus subtilis. *J. Bacteriol.* **193**, 4821–4831. (doi:10.1128/JB.00223-11)

19. Kobayashi, K. & Iwano, M. 2012 BslA(YuaB) forms a hydrophobic layer on the surface of Bacillus subtilis biofilms. *Mol. Microbiol.* **85**, 51–66. (doi:10.1111/j.1365-2958.2012.08094.x)

20. Hobley, L., Ostrowski, A., Rao, F. V., Bromley, K. M., Porter, M., Prescott, A. R., Macphee, C. E., van Aalten, D. M. F. & Stanley-Wall, N. R. 2013 BslA is a self-assembling bacterial hydrophobin that coats the Bacillus subtilis biofilm. *Proc. Natl. Acad. Sci. U. S. A.* **110**, 13600–13605. (doi:10.1073/pnas.1306390110)

21. Bromley, K. M. et al. 2015 Interfacial self-assembly of a bacterial hydrophobin. *Proc. Natl. Acad. Sci. U. S. A.* **112**, 5419–5424. (doi:10.1073/pnas.1419016112)

22. Morris, R. J., Bromley, K. M., Stanley-Wall, N. R. & MacPhee, C. E. 2016 A phenomenological description of BslA assemblies across multiple length scales. *Philos. Trans. A* **374**, 20150131. (doi:http://dx.doi.org/10.1098/rsta.2015.0131)

23. Schloss, A. C. et al. 2016 Fabrication of Modularly Functionalizable Microcapsules Using Protein-Based Technologies. *ACS Biomater. Sci. Eng.* **2**, 1856–1861. (doi:10.1021/acsbiomaterials.6b00447)

24. Tcholakova, S., Denkov, N. D. & Lips, A. 2008 Comparison of solid particles, globular proteins and surfactants as emulsifiers. *Phys. Chem. Chem. Phys.* **10**, 1608–1627. (doi:10.1039/b715933c)

25. Walstra, P. & Smulders, P. E. A. 1998 Emulsion Formation. In *Modern Aspects of Emulsion Science* (ed. B. P. Binks), pp. 56–99. Cambridge: The Royal Society of Chemistry.

26. Tjahjadi, M., Stone, H. A. & Ottino, J. M. 1992 Satellite and subsatellite formation in capillary breakup. *J. Fluid Mech.* **243**, 297–317. (doi:10.1017/S0022112092002738)

27. French, D. J., Taylor, P., Fowler, J. & Clegg, P. S. 2015 Making and breaking bridges in a Pickering emulsion. *J. Colloid Interface Sci.* **441**, 30–38. (doi:10.1016/j.jcis.2014.11.032)

28. Cristini, V., Guido, S., Alfani, A., Błwzdziewicz, J. & Loewenberg, M. 2003 Drop

breakup and fragment size distribution in shear flow. *J. Rheol.* **47**, 1283-1298. (doi:10.1122/1.1603240)

29. Swift, D. L. & Friedlander, S. 1964 The coagulation of hydrosols by brownian motion and laminar shear flow. *J. Colloid Sci.* **19**, 621–647. (doi:10.1016/0095-8522(64)90085-6)

30. Studart, A. R., Shum, H. C. & Weitz, D. A. 2009 Arrested coalescence of particle-coated droplets into nonspherical supracolloidal structures. *J. Phys. Chem. B* **113**, 3914–3919. (doi:10.1021/jp806795c)

31. Santos, H. M., Lodeiro, C. & Capelo-Martínez, J.-L. 2008 The Power of Ultrasound. In *Ultrasound in Chemistry: Analytical Applications* (ed. J.-L. Capelo-Martínez), pp. 1–15. Weinheim: Wiley-VCH.

32. Weninger, K. R., Barber, B. P. & Putterman, S. J. 1997 Pulsed Mie scattering measurements of the collapse of a sonoluminescing bubble. *Phys. Rev. Lett.* **78**, 1799–1802. (doi:10.1103/PhysRevLett.78.1799)

33. Xu, S. et al. 2005 Generation of monodisperse particles by using microfluidics: Control over size, shape, and composition. *Angew. Chemie - Int. Ed.* **44**, 724–728. (doi:10.1002/anie.200462226)

34. Prileszky, T. A. & Furst, E. M. 2016 Fluid networks assembled from endoskeletal droplets. *Chem. Mater.* **28**, 3734–3740. (doi:10.1021/acs.chemmater.6b00497)